\documentclass[prb,reprint,aps]{revtex4-1} 
\usepackage[utf8]{inputenc}
\usepackage[fleqn]{amsmath}
\usepackage{amssymb}
\usepackage{amsmath}
\usepackage{url}
\usepackage{natbib}
\usepackage{hyperref}
\usepackage[margin=1in]{geometry}
\usepackage{graphicx} 
\usepackage{rotating}

\usepackage{footmisc}
\DefineFNsymbols{mySymbols}{{\ensuremath\dagger}{\ensuremath\ddagger}\S\P
   *{**}{\ensuremath{\dagger\dagger}}{\ensuremath{\ddagger\ddagger}}}
\setfnsymbol{mySymbols}

\begin{document}

\title{Plane Pendulum and Beyond by Phase Space Geometry}

\author{Bradley Klee}
\email{bjklee@email.uark.edu, bradklee@gmail.com} 
\affiliation{Department of Physics, University of Arkansas, Fayetteville, AR 72701}

\date{\today}

\begin{abstract}
The small angle approximation often fails to explain experimental data, does 
not even predict if a plane pendulum's period increases or decreases with increasing 
amplitude. We make a perturbation ansatz for the Conserved Energy Surfaces of a one-dimensional, 
parity-symmetric, anharmonic oscillator. A simple, novel algorithm produces the equations of motion 
and the period of oscillation to arbitrary precision. The Jacobian elliptic functions appear 
as a special case. Thrift experiment combined with recursive data analysis provides experimental 
verification of well-known predictions. Development of the quantum/classical analogy 
enables comparison of time-independent perturbation theories. Many of the useful notions herein 
generalize to integrable and non-integrable systems in higher dimensions.  
\end{abstract}

\maketitle 

\section{Introduction}

\begin{figure}[h] 
\begin{center}
\includegraphics[scale=.45]{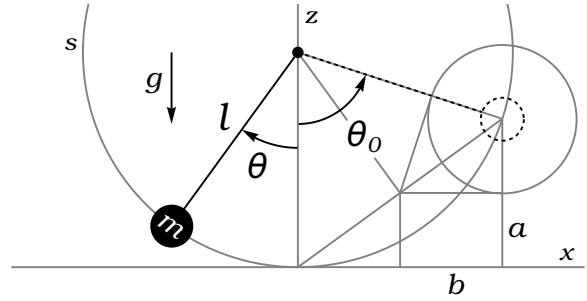} 
  \caption{Simple pendulum and coordinate geometry. Half height $a$ determines the period of oscillation.}
  \label{fig:CoordGeo}
\end{center}
\end{figure}

Space and time play foundational roles in all experiments and most 
equations. Measurement of \textit{space} only requires the definition of a standard 
length. What difficulties prevent easy measurement of time? A day or year is too long for 
describing the fall of an apple, while heartbeats depend on unpredictable biological 
conditions. A pendulum, as in Fig.~\ref{fig:CoordGeo}, oscillates through 
\textit{time}, setting a scale on the order of one second when 
$l \approx 1/4(m)$ and $g \approx 9.81(m/s^2)$.

History credits Galileo with early discoveries regarding the time dependent behavior 
of oscillating pendulums \cite{Ariotti}. He noticed the \textit{isochrony} of identical 
pendulums, manifest as one characteristic time, the period. Galileo's initial pronouncement 
that the period depends not on amplitude now resounds false. The flourishing of classical 
mechanics provides a logical alternative in perturbation theory. Experiments yielding digital 
data support predictions contrary to the musings of Galilean renaissance. 

The timing of a pendulum does depend on its initial condition\cite{Landau,HarterBook}. 
After Legendre and Abel, C.G.J. Jacobi standardized and optimized the solution of the pendulum's motion by 
defining a set of elliptic functions\cite{Jacobi,Brizard}. The Jacobian elliptic functions 
have many interpretations in physics, but do not fall into the core curriculum 
because they present serious technical challenges \cite{JacobiBead,JacobiErdos}. 
As an alternative the physics literature contains a variety of approximate solution methods, including: 
\textit{ad hoc} \cite{MeanApproximation,KFApproximation},  Lindstedt-Poincar\'{e} 
\cite{Landau,Fulcher,Park}, and canonical perturbation theory\cite{Lowenstein}.

Our novel method views the pendulum as a one-dimensional, anharmonic oscillator with parity 
symmetry. In a two-dimensional phase space spanned by position and momentum coordinates, 
geometric methods apply. Motion happens along a Conserved Energy Surface (C.E.S.), which 
is not too different from the perimeter of a circle. We make a deformation ansatz and apply 
an iterative algorithm that sets undetermined functions to force convergence of the energy 
to one conserved value. One dimensional oscillators are integrable systems, so time dependence 
follows readily. 

The program of derivation involves no mistaken assumptions and refuses temptation to plagiarize 
standard references. By solving the pendulum equations of motion to high precision, we obtain a 
series expansion of the Jacobi Elliptic functions $sn(\vartheta,\alpha)$ and $cn(\vartheta,\alpha)$. 
This exercise distinguishes the derivation as arbitrary-precision and free of error. 

A third approximation suffices to describe the pendulum's motion through one period so long 
as the motion obeys $\theta \in [-\pi/2,\pi/2]$. A thrift experiment takes place in this range. 
The setup utilizes a modified USB mouse to produce digital data. Analysis yields extracted 
parameters that closely agree with carefully derived predictions.

Ultimately we reveal the details of a semiclassical analogy between time-independent perturbation 
theories. In quantum mechanics approximate wavefunctions must nearly conserve energy. Comparing 
results for quartic oscillators, we derive a quantum condition, which is equivalent to the 
Sommerfeld-Wilson prescription in the classical limit.    

\section{Dimensional Analysis}
The plane pendulum consists of a massive bob attached by a string 
or a rod, assumed massless, to an axle as in Fig.~\ref{fig:CoordGeo}\cite{HarterBook}. 
Gravity acts on the bob with vertical force $mg$, and the attachment applies a response 
force, the tension. As time elapses the bob swings and executes a periodic motion 
along a circular trajectory of radius $l$. In \textit{librational motion}
the sign of $\dot{\theta}$ alternates while the pendulum reaches a maximum 
deflection $\pm\theta_0$ at regular intervals throughout the experiment. The 
time of one complete oscillation, say from $\theta_0$ to $-\theta_0$ back to $\theta_0$, is called the 
\textit{period}. 

Table \ref{tab:PQList} collects the relevant physical quantities, read directly 
from Fig.~\ref{fig:CoordGeo}. Dimensional symbols $[\;L\;]$, $[\;M\;]$, and $[\;T\;]$, denote 
length, mass, and time. The quantities $\{\sqrt{l/g},\sqrt{a/g},\sqrt{b/g}\}$ all have 
dimension of time, $[\;T\;]$. Assuming Galileo's observation correct, $\sqrt{l/g}$ must 
be the dimensional scale of time because quantities $\{a,b\}$ depend on the amplitude 
of motion. 

\begin{table}[h]
\begin{center}
\caption{Dimensional Quantities.} 
\label{tab:PQList}
\begin{tabular}{ c | c | c  }
\hline \hline
Symbol & \;\;\;\; Dimension \;\;\;\; & Trigonometric Form  \\
\hline
$l$ & $[\;L\;]$ & $\cdot$ \\
$a$ & $[\;L\;]$ & $l\;\sin^2(\theta_0/2)$ \\
$b$ & $[\;L\;]$ & $\;\;l\;\sin(\theta_0/2)\;\cos(\theta_0/2)\;\;$ \\
$g$ & $[\;L\;]\;[\;T\;]^{-2}$ & $\cdot$ \\
$m$ & $[\;M\;]$ & $\cdot$ 
\end{tabular}
\end{center}
\end{table}

A naive energy argument improves the estimation. The maximum potential energy is 
$2 m g a$. Assume that this energy converts entirely to kinetic energy
as the mass $m$ moves at constant velocity $2\sqrt{ga}$ through a distance $8b$, then
$T_0 \approx 4b/\sqrt{ga} $. In the small angle approximation, $a \ll l$, 
$b \approx \sqrt{l\;a}$, and $T_0 \approx 4\;\sqrt{l/g} $, certainly 
an underestimate. 

The exact period follows from a more sophisticated calculation, again based on conservation 
of energy. At any half-height $z=(l/2)(1-\cos(\theta))<a$, the kinetic energy equals $2mg(a-z)$, the velocity 
equals $2\sqrt{g(a-z)}$, and the period equals\cite{Landau,Brizard}
\begin{eqnarray}
T(a/l) &= \int_0^{T}dt =  4\int_0^{l\theta_0}\frac{ds}{2\sqrt{g(a-z)}} \\
 &=   4 \sqrt{\frac{l}{g}}\int_0^{\pi/2}\frac{d\xi}{\sqrt{1-(a/l)\sin^2(\xi)}}
 =   4 \sqrt{\frac{l}{g}} K(\frac{a}{l}),\nonumber
\end{eqnarray}
where $ds$ goes along the arc of motion. The complete elliptic integral of the first kind, $K$, 
admits no simple closed-form. Alternatively, the small angle approximation 
eliminates dependence on $a$ and gives a concise result 
\begin{eqnarray}
T_0 &= \lim\limits_{a\rightarrow 0}  4\int_0^{2\;b}\frac{dx}{2\sqrt{g(a-z)}} \\
 &= \lim\limits_{a\rightarrow 0}  2\sqrt{\frac{l}{g}}\int_0^{a}\frac{dz}{\sqrt{z(a-z)}} =2\pi\sqrt{\frac{l}{g}} ,  \nonumber 
\end{eqnarray}
which requires small-angle identity $s \approx x \approx  2\sqrt{lz}$ to change from the circular line element $ds$ to 
the horizontal $dx$, and finally to the vertical $dz$. The simple result only applies 
in the limit $a\rightarrow0$. 

In a general one-dimensional oscillation with small-amplitude period $T_0$,
we usually have something along the lines 
\begin{equation}
T(\alpha,\boldsymbol\epsilon) = f(\alpha,\boldsymbol\epsilon)\; T_0  ,
\end{equation}
with $f(\alpha,\boldsymbol\epsilon)$ a complicated function 
of dimensionless energy $\alpha$, and $\boldsymbol\epsilon$, structure 
constants of the potential energy. 

With the pendulum experiment the trouble is in the initial conditions. Each initial condition 
determines one critical parameter 
\begin{equation}
\alpha = a/l = \frac{1}{2}\big( 1 - \cos(\theta_0) \big) =  \sin^2(\theta_0 / 2), 
\end{equation}
proportional to the total energy. In the simple harmonic 
approximation $\alpha$ tends to zero as $\theta_0^2$. Considering this fact, 
the hypothesis that factor $f(\alpha,\boldsymbol\epsilon) \rightarrow f(\alpha)$ 
has a non-terminating power series expansion seems likely. Referencing the expansion 
of $K(\alpha)$ \cite{Wolfram} we have
\begin{eqnarray}
 f(\alpha)  =\frac{T(\alpha)}{T_0} = \frac{2}{\pi}K(\alpha) 
 = \sum_{n=0}^{\infty} \bigg(\frac{(2\;n-1)!!}{(2\;n)!!}\bigg)^2\;\alpha^n  \nonumber\\
  = 1+\frac{\alpha}{4} + \frac{9 \alpha^2}{64} + \frac{25 \alpha^3}{256} \ldots 
  \;\;\;\;\;\;\;\;\;\;\;\;\;\;\;\;\;\;\;\;\;\; 
\end{eqnarray}
 
Numerous approximation schemes (cf.  \cite{MeanApproximation}, Table III) aim to simplify the
description of a pendulum's \textit{anharmonicity}, as measured by coefficients to the 
powers of $\alpha$. At small $\alpha$ all formulae for $T(\alpha)$ must asymptotically approach Eq.5, so 
exact and approximate agreement to $\mathcal{O}(\alpha)$ and $\mathcal{O}(\alpha^2)$ 
respectively is a common feature of the many published results. For example, 
the empirical Kidd-Fogg formula \cite{KFApproximation} has
\begin{equation}
f(\alpha) =\frac{1}{(1-\alpha)^{1/4}} =  1+\frac{\alpha}{4} + \frac{5 \alpha^2}{32} + \frac{15 \alpha^3}{128} \ldots 
\end{equation}

Of course, other factors introduce uncertainty to physical experiments \cite{Nelson}, and 
these uncertainties always cause the data to deviate from theoretical expectations. Say that we 
write the standard deviation $\sigma$ in units of $T_0$, then $\sigma$ competes in order of 
magnitude with terms from the expansion of $f(\alpha)$ until eventually, for some integer $n$, we have $c_n\;\alpha^n \ll \sigma$. 
This logic is useful in data analysis and gives some restraint to our exploration of approximate solutions. 

\section{Phase Space Geometry}

\subsection{Small Angle Approximation}
 \begin{figure}[h] 
\begin{center}
\includegraphics[scale=.5]{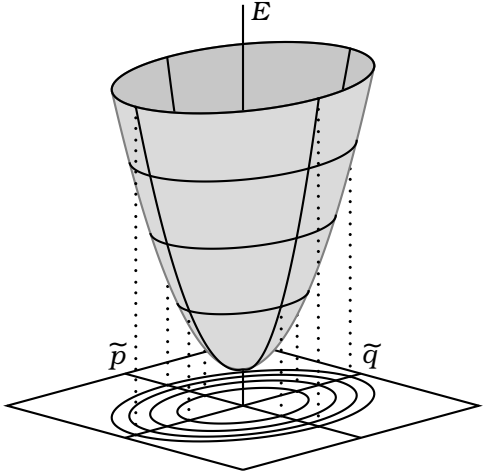} 
  \caption{Total Energy Surface. Level sets of the total energy function project 
  trajectories into the plane of phase space, the Conserved Energy Surfaces.   }
  \label{fig:Para}
\end{center}
\end{figure} 
In terms of the phase space coordinates, $(\widetilde{q},\widetilde{p})=(l\;\theta , m\;l\;\dot{\theta})$,
the pendulum kinetic and potential energy are 
\begin{equation}
T = \widetilde{p}^2/2m, \;\;\;\;\; V = m \;g\; l \;(1-\cos(\widetilde{q}/l)).
\end{equation}
The potential expands in power series 
\begin{equation}
V = m \;g\; l \;\sum_{n=1} \frac{(-1)^{n+1}}{(2n)!}(\widetilde{q}/l)^{2n}.
\end{equation}
In the small angle approximation we assume that $\widetilde{q} \ll l$ throughout the experiment. 
Keeping only the first potential term allows us to write the conserved, total energy of the 
pendulum oscillator in the small angle approximation
\begin{equation}
E(\alpha) = 2\;m\;g\;l\;\alpha \approx  \frac{1}{2}(\frac{1}{m}\widetilde{p}^2 + \frac{m\;g}{l} \widetilde{q}^2).
\end{equation}
Clearly there exists a bijection between energies and elliptical 
trajectories, depicted as a projection in Fig.~\ref{fig:Para}. 
Define radius $\Psi$ and angle $\phi$ the polar coordinates of phase space. 
Each closed curve $\Psi(\alpha,\phi) \rightarrow \Psi(\alpha)$ is alternatively a 
\textit{phase space trajectory } or a \textit{Conserved Energy Surface} (C.E.S.). 

It is much easier to determine time dynamics in a system of measurement where the phase 
space trajectory takes the particular form of a circle, so we need to apply a canonical 
transformation\cite{RalstonSymp}
\begin{subequations}
\begin{align}
 \widetilde{q} \rightarrow q   & = \bigg(\frac{m^2\;g}{l}\bigg)^{1/4}\;\widetilde{q}\;, \\
 \widetilde{p} \rightarrow p  & = \bigg(\frac{l}{m^2\;g}\bigg)^{1/4}\;\widetilde{p}\;, \\
 E(\alpha) \rightarrow E(\alpha) & =  \frac{\omega_0}{2}\big(p^2 + q^2\big),
\end{align}
\end{subequations}
with $\omega_0=\sqrt{g/l}$. Transformation Eq. 10 takes elliptical trajectories into circular trajectories 
with equal energy and equal enclosed \textit{phase area}
\begin{equation}
 \lambda(\alpha) = \oint d\widetilde{q} \; \widetilde{p}(\alpha,\widetilde{q}) = \oint dq \; p(\alpha,q).
\end{equation}
The period
\begin{equation}
T(\alpha) = \oint dt = \oint d\widetilde{q}\;\frac{m}{\widetilde{p}(\alpha,\widetilde{q})} = \oint \frac{dq}{\omega_0\;p(\alpha,q)}, 
\end{equation}
also remains invariant under the canonical transformation Eq. 10. To see this we use another 
definition\cite{Arnold} for the period
\begin{equation}
T = \frac{d\lambda}{dE} = \oint dq \bigg(\frac{dE}{dp}\bigg)^{-1} = \oint \frac{dq}{\omega_0\;p} , 
\end{equation}
with $\alpha$ dependence suppressed. This equation  proves a connection between the physical period of motion 
and the purely geometric phase area. 
As $\lambda(\alpha)$ and $E(\alpha)$ remain invariant under canonical transformation
so too must $T(\alpha)$. 

Circular trajectories transform invariantly under rotations around the origin of phase 
space, which immediately implies $\ddot{\phi}=0$. Then the time-dependent 
solution to the equations of motion is
\begin{subequations}
\begin{align}
 q = \Psi(\alpha,0) \; \cos\big(-\omega_0 ( t-t_0) \big) , \\
 p = \Psi(\alpha,0) \; \sin\big(-\omega_0 ( t-t_0) \big) ,
\end{align}
\end{subequations}
with angular frequency $\omega_0 = 2\;\pi/T_0$, $t_0$ an arbitrary constant. 

The small angle approximation does not say anything about the expansion for $T(\alpha)$. To 
illustrate the dangers of approximation, let us work out a clever ruse. With 
$\widetilde{q} \ll l \Longrightarrow l\theta_0 \approx 2b$ we have 
\begin{eqnarray}
\Psi(\alpha,0)^2 & = (\frac{m^2 g}{l})^{\frac{1}{2}}(2b)^2= 4\;\beta^2\;m \;\sqrt{g\;l^3} \;\;\;\; \\ 
              & = 4\;\alpha\;(1-\alpha)\;m\;\sqrt{g\;l^3} \nonumber,
\end{eqnarray}
where $\beta^2 = (b/l)^2 = \alpha\;(1-\alpha)$. The constant radius $\Psi(\alpha,0)$ determines 
the phase area bounded by closed-curve $\Psi(\alpha)$, again assumed circular,
\begin{eqnarray}
\lambda(\alpha) &= \oint p(\alpha,q) \;dq = \int_0^{2\;\pi} d\phi \int_0^{\Psi(\alpha,0)} r \;dr \;\;\;\;\; \\
& = \pi\;\Psi(\alpha,0)^2 = \lambda_0\;(\alpha-\alpha^2), \nonumber
\end{eqnarray}
where $\lambda_0 = 2\;m\;g\;l\;T_0\;$. We have yet to determine $T(\alpha)$, and do 
not assume that $T(\alpha)=T_0$. Instead we calculate $T(\alpha)$ by the beautiful 
formula Eq. 13. 

\begin{figure}[h] 
\begin{center}
\includegraphics[scale=.35]{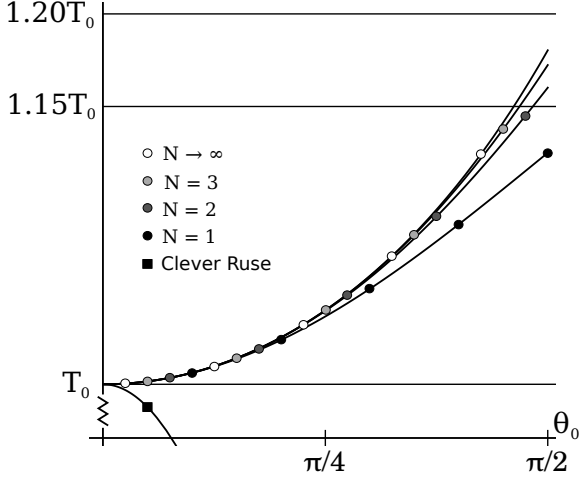} 
  \caption{Comparison of Period Approximations. The small angle approximation doesn't prevent wrong, divergent predictions.
	   Labels for convergent approximations follow naming convention of section IIIB. }
  \label{fig:PeriodGraph}
\end{center}
\end{figure}

Our \textit{phase space geometry} consists of a triple 
$\{E, \Psi , \lambda \}$. In the small angle approximation, 
height $E(\alpha)$ is an exact function of $\alpha$ while perimeter 
$\Psi(\alpha)$ and area $\lambda(\alpha)$ are merely approximations, 
so we expect to find inconsistency in the geometry wherever the 
assumptions break down,  
\begin{equation}
f(\alpha) = \frac{1}{T_0} \frac{d \alpha}{d E} \frac{d \lambda}{d\alpha} = \frac{1}{\lambda_0} \frac{d \lambda}{d\alpha} = (1-2\;\alpha).
\end{equation}
Comparing with Eq. 5, we see that $T(\alpha)$ in the small angle approximation  
may give the wrong $\mathcal{O}(\alpha)$ asymptote as depicted in 
Fig.~\ref{fig:PeriodGraph}. Worse, the small angle approximation 
allows us to predict incorrectly that the period \textit{decreases} 
with increasing total energy! 

\subsection{Simple Anharmonic Approximation}

The \textit{reductio ad absurdum} of section III.A clearly states the need to find a better approximation 
of the exact phase space geometry. To present results in a more general fashion, we 
treat the pendulum as an anharmonic oscillator with a potential $V$. The potential expands in power series  
around a position of stable equilibrium, i.e., 
$\frac{\partial V}{\partial q}|_{q=0} = 0,\frac{\partial^2 V}{\partial q^2}|_{q=0}>0$. 
Imposing the symmetry constraint $V(q) = V(-q)$, the most general form for the total energy
reduces to
\begin{equation}
E  =\frac{\omega_0}{2}\big(p^2 + q^2\big) +\sum_{n=1}\frac{\omega_0}{(2(n+1))!}\frac{\epsilon_{n}}{\lambda_{\pi}^{n}}q^{2(n+1)}, 
\end{equation}
We make an ansatz of the form
\begin{equation}
 \Psi(\alpha,\phi) = \sqrt{2\;\lambda_{\pi}\;\alpha} \; \bigg(\;1 + \sum_n  \alpha^n \; \psi_n(\phi)\;\bigg) ,\;\;\;\;\; 
\end{equation}
with $\lambda_{\pi} = \frac{\lambda_0}{2\;\pi}$. 

Our strategy is to substitute $\Psi(\alpha,\phi)$ into the energy equation, and determine the functions $\psi_n(\phi)$ in terms of the 
expansion coefficients $\epsilon_n$ by forcing the energy to equal $\lambda_{\pi}\omega_0\alpha + \mathcal{O}(\alpha^{N+2})$ for 
some integer $N \ge 0$. As the C.E.S. more nearly obeys conservation of energy, the approximation improves. 

\subsubsection{$N=1$, The $\mathcal{O}(\alpha)$ Approximation}
A first approximation only requires the first term of each sum in Eqs. 18 \& 19. Applying 
$(q,p)\rightarrow(\Psi(\alpha,\phi)\cos(\phi),\Psi(\alpha,\phi)\sin(\phi))$ to the energy 
equation and collecting terms by order, we have
\begin{subequations}
 \begin{align}
\alpha \;&:\;\lambda_{\pi}\omega_0\alpha, \\
\alpha^2 \;&:\; \lambda_{\pi} \omega_0 \alpha^2 \big(\frac{\epsilon_1}{6}\;\cos(\phi)^4+2\;\psi_1(\phi)\big).
\end{align}
\end{subequations}
Setting terms at $\mathcal{O}(\alpha^2)$ equal to zero and solving for $\psi_1(\phi)$, 
we find
\begin{equation}
 \Psi(\alpha,\phi) = \sqrt{2\;\lambda_{\pi}\;\alpha}\big(1-\frac{\epsilon_1}{12}\;\alpha\;\cos^4(\phi)+\mathcal{O}(\alpha^2)\big).
\end{equation}
As in section III.A. the phase space geometry determines approximate quantities 
\begin{subequations}
 \begin{align}
\lambda(\alpha) &= \oint p(\alpha,q) \;dq = \int_0^{2\;\pi} d\phi \int_0^{\psi(\alpha,\phi)} r \;dr \;\;\;\;\; \nonumber \\
& =  \lambda_0\;(\;\alpha-\frac{\epsilon_1\;\alpha^2}{16}) + \mathcal{O}(\alpha^3),  \\
f(\alpha) & = 1-\frac{\epsilon_1\;\alpha}{8} + \mathcal{O}(\alpha^2). \;\;\;\; 
\end{align}
\end{subequations}
The pendulum has $\epsilon_1=-2$, which makes Eq. 22b asymptotic with Eq. 5 to 
$\mathcal{O}(\alpha)$. 

\subsubsection{$N=2$, The $\mathcal{O}(\alpha^2)$ Approximation}
The approximation improves if we include summands for $n=1$ and $n=2$.
Evaluation of the energy yields the same constraints as above and 
\begin{eqnarray}
\alpha^3 \;&:\; \lambda_{\pi} \omega_0 \alpha^3\big(\;
\frac{1}{90}\; \epsilon_2\;\cos^6(\phi)  + 2 \; \psi_2(\phi)
 \\
&+ \frac{2}{3} \;\epsilon_1\;\cos^4(\phi) \; \psi_1(\phi) +   \psi_1(\phi)^2\;\big). \nonumber
\end{eqnarray}
Setting $\mathcal{O}(\alpha^3)$ terms equal to zero, 
substituting the determined form of $\psi_1(\phi)$,
and solving for $\psi_2(\phi)$ determines
\begin{eqnarray}
 \Psi(\alpha,\phi) & = \sqrt{2\;\lambda_{\pi}\;\alpha}\big(1-\frac{\epsilon_1}{12}\;\alpha\;\cos^4(\phi) \\
 & +\frac{7\;\epsilon_1^2}{288}\;\alpha^2\;\cos^8(\phi) -\frac{\epsilon_2}{180}\;\alpha^2\;\cos^6(\phi)+\mathcal{O}(\alpha^3)\big). \nonumber
\end{eqnarray}
The estimation of $f(\alpha)$ slightly improves,
\begin{subequations}
 \begin{align}
\lambda(\alpha) &= \lambda_0\;(\alpha-\frac{\epsilon_1\;\alpha^2}{16} \\
&+\frac{35\;\epsilon_1^2 \;\alpha^3}{2304}-\frac{\epsilon_2\;\alpha^3}{288}) + \mathcal{O}(\alpha^4) \nonumber \\
f(\alpha) & = 1  - \frac{\epsilon_1\;\alpha}{8} \\
& + \frac{35\;\epsilon_1^2 \;\alpha^2}{768}-\frac{\epsilon_2\;\alpha^2}{96}
+ \mathcal{O}(\alpha^3). \nonumber
\end{align}
\end{subequations}
Inserting pendulum values $(\epsilon_1,\epsilon_2)=(-2,4)$ makes Eq. 25b 
asymptotic with Eq. 5 to $\mathcal{O}(\alpha^2)$.

\subsubsection{The $\mathcal{O}(\alpha^N)$ Approximation}

By iterating the procedure applied for $N=1$ and $N=2$, we obtain an approximation 
to arbitrary order. Every $\psi_n(\phi)$ can be expanded in Fourier series or in 
even powers of cosine. For smooth potentials with a single minimum, the 
approximation converges according to  
\begin{equation}
\lim_{N \rightarrow \infty}E=\lim_{N \rightarrow \infty} \lambda_{\pi}\omega_0\alpha 
+ \mathcal{O}(\alpha^{N+2}) = \lambda_{\pi}\omega_0\alpha.
\end{equation}
If the $\boldsymbol\epsilon$ coefficients grow rapidly or contain a divergence, then more 
detailed analysis is required. 

A simple symbolic computation calculates higher order expansions by routine. Taking the pendulum as 
an example with $\epsilon_n = (-2)^n$, we write a simple code, and store expansion coefficients in 
the Online Encyclopedia of Integer Sequences \cite{OEIS} (Cf. 
\href{https://oeis.org/A273506}{A273506},
\href{https://oeis.org/A273507}{A273507},
\href{https://oeis.org/A274130}{A274130},
\href{https://oeis.org/A274131}{A274131},
\href{https://oeis.org/A274076}{A274076},
\href{https://oeis.org/A274078}{A274078}). Relaxing the condition $V(q)=V(-q)$, we also calculate 
various expansions for a potential where the $\boldsymbol\epsilon$ variables take on arbitrary values 
(Cf. 
\href{https://oeis.org/A276738}{A276738},
\href{https://oeis.org/A276814}{A276814},
\href{https://oeis.org/A276815}{A276815},
\href{https://oeis.org/A276816}{A276816}).

\textit{Mathematica} algorithms available via OEIS entries \href{https://oeis.org/A273506}{A273506} and 
\href{https://oeis.org/A276816}{A276816} give two different ways to compute 
arbitrary precision expansions of phase space trajectories and $K(\alpha)$\cite{Klee}. Whenever $N<10$, 
these algorithms operate in small time on a personal computer. For moderate ranges of $\alpha$, enumeration 
beyond $N=3$ follows a law of diminishing returns. As can be seen in Fig.~\ref{fig:PeriodGraph}, the 
$\mathcal{O}(\alpha^3)$ approximation already captures to within $1\%$, the exact behavior of the pendulum 
in the range $\alpha \in [0,1/2]$,  $\theta_0 \in [-\pi/2,\pi/2]$, where our experiment takes place. 

\subsection{Time Dependence}
The phase space trajectory determines time evolution
\begin{eqnarray}
dt = \frac{dq}{\omega_0\;p} = \frac{d\phi}{\omega_0} \bigg(\frac{\Psi'(\alpha,\phi)}{\Psi(\alpha,\phi)}\; cot(\phi)-1 \bigg),
\end{eqnarray}
where prime indicates differentiation with respect to $\phi$. Again expand in powers of 
$\alpha$
\begin{eqnarray}
 \frac{\Psi'(\alpha,\phi)}{\Psi(\alpha,\phi)} =  \alpha \; \psi_1'(\phi) 
 + \alpha^2\;\big( \; \psi_2'(\phi) - \\
 \psi_1(\phi)\;\psi_1'(\phi)\;\big) +\mathcal{O}(\alpha^3), \nonumber
\end{eqnarray}

Between two near points in phase space 
\begin{eqnarray}
 dt &  \approx  \frac{d \phi}{\omega_0}\big(-1+ \frac{1}{3} \; \cos^4(\phi)\;\epsilon_1\;\alpha \\
 & + (\;\frac{1}{30}\;\cos^6(\phi)\;\epsilon_2 -\frac{1}{6}\;\cos^8(\phi)\;\epsilon_1^2\;)\;\alpha^2 \; \big),  \nonumber
\end{eqnarray}
where we drop terms higher than $\mathcal{O}(\alpha^2)$. Expanding cosine terms (Cf. \href{https://oeis.org/A273496}{A273496}) allows direct integration 
of $dt$; however, results at high order are not easy to express in concise form. To first order 
\begin{eqnarray}
t_1 &= \int_0^{t_1} dt = -\frac{\phi_1}{\omega_0}(1-\frac{\epsilon_1\;\alpha}{8})  \\ 
&+\frac{\alpha\;\epsilon_1}{\omega_0}\big(\frac{1}{12} \sin(2\phi_1) + \frac{1}{96} \sin(4\phi_1) \big) + \mathcal{O}(\alpha^2), \nonumber
\end{eqnarray}
with limits $\phi(0)=0$ and $ \phi(t_1)=\phi_1$. By Lagrange inversion we could in principle obtain 
$\phi_1(t_1)$ \cite{Lang,WW27}.

Alternatively, we have 
\begin{eqnarray}
\frac{d\phi}{dt} =  \dot{\phi}(\phi) = \cos^{2}(\phi)\frac{d}{dt}\bigg( \tan(\phi)\bigg).
\end{eqnarray}
Using the equations of motion and substituting an approximation for 
$\frac{d}{dt}\tan(\phi)=(q\dot{p}-\dot{q}p)/q^2$, we obtain expressions for the phase 
space angular velocity, such as
\begin{eqnarray}
\frac{d\phi}{dt} \approx -\omega\bigg( \; 1+ \frac{1}{3} \; \cos^4(\phi)\;\epsilon_1\;\alpha 
\;\;\;\;\;\;\;\;\;\;\;\;\;\;\;\;\;\;\;\;\; \\   
  + \big(\frac{1}{30}\; \cos^6(\phi) \; \epsilon_2 -\frac{1}{18} \; \cos^8(\phi)\; \epsilon_1^2 \; \big)\; \alpha^2 \; \bigg).\nonumber
\end{eqnarray}
The phase velocity $\dot{\phi}$ depends on the phase angle $\phi$, as expected in 
any oscillation where the phase space trajectory deforms away from elliptical or circular 
shape. Either limit $(\epsilon_1,\epsilon_2) \rightarrow (0,0)$ or $\alpha \rightarrow 0$ recovers 
constant $\dot{\phi}$, when again trajectories become ellipses or circles. 

\begin{figure}[h] 
\begin{center}
\includegraphics[scale=.28]{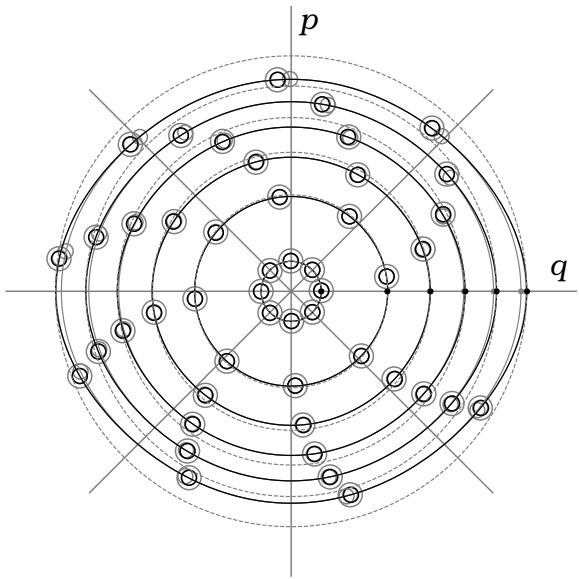} 
  \caption{Time Evolution of Pendulums in Phase Space.  The $N=10$ ( black ), $N=1$ ( gray ), 
  and circular ( dashed gray ) trajectories are plotted for $ \alpha = \{0.01,0.1,0.2,0.3,0.4,0.5 \} $. 
  The $N=10$ trajectory completely overlaps the $N=3$ trajectory for all $\alpha$ values, showing 
  convergence. Small filed circles mark initial conditions, while open circles indicate the state of a system 
  at intervals of $\Delta t = T_0/8$ as it rotates clock-wise through phase space. Large open circles are 
  calculated by the technique of symplectic integration.}
  \label{fig:TimeEvo}
\end{center}
\end{figure}

There are numerous well known methods for calculating numerical time evolution of a Hamiltonian 
system including symplectic integration \cite{Symplectic}. In the present case, time evolution 
occurs along the one-dimensional C.E.S. The standard Runge-Kutta algorithm applies; though, 
the task of integration requires only minimal complexity. The simple Euler's method (RK1) suffices. 
Iteration through time according to $\dot{\phi}$ yields time-dependent predictions, as depicted 
in Fig.~\ref{fig:TimeEvo}. This is the first plot to clearly show anharmonicity as anisochronous 
motion of pendulums with different initial conditions. 

Around $\alpha = 0$ all approximations become indistinguishable. To $\alpha=1/2$, the  
$\mathcal{O}(\alpha)$ approximation closely matches the exact numerical solution, which 
is nearly indistinguishable from the $\mathcal{O}(\alpha^3)$ approximation. Isochrony becomes 
more pronounced at high $\alpha$ where, after $8$ intervals of $\Delta t = T_0/8$, 
the pendulum does not nearly reach its initial condition. 

\section{Comparison with Standards}

\subsection{Jacobian Elliptic Functions}
The Jacobian elliptic functions determine exactly the phase space geometry of the simple pendulum. 
The properties of these functions are well known and recorded in a number of standard resources
\cite{Abramowitz,WW27,Wolfram}. Paul Erd\"os gives a creative, geometric introduction via the 
Seiffert spirals\cite{JacobiErdos}.

The exact pendulum phase space trajectory is 
\begin{subequations}
 \begin{align}
q(\alpha,\vartheta) &= \sqrt{2 \;\lambda_{\pi} }\;\arcsin(\sqrt{\alpha}\;sn(\vartheta,\alpha)) , \\
p(\alpha,\vartheta) & =  
\sqrt{2 \;\lambda_{\pi}\;\alpha}\;cn(\vartheta,\alpha)  , \\
\Psi(\alpha,\vartheta) &=\sqrt{q(\alpha,\vartheta)^2+p(\alpha,\vartheta)^2},
\end{align}
\end{subequations}
where $cn$, $sn$ are Jacobian elliptic functions of angular coordinate $\vartheta = K(\alpha)+\omega_0\;t$ 
with period $4K(\alpha)$. 

Substituting in time dependence such as Eq. 30, it is possible to expand $\Psi(\alpha,\vartheta)$ in 
powers of $\alpha$ and prove, order-by-order, equivalence between the exact solution and the 
approximate solution of III.B. We need not perform this tedious calculation, for any  
solution that conserves energy must be equivalent to the exact solution. Rather, let us explore 
convergence by plotting the approximations of $sn$ and $cn$ near the divergence $\alpha=1$. 

\begin{figure}[h] 
\begin{center}
\includegraphics[scale=.5]{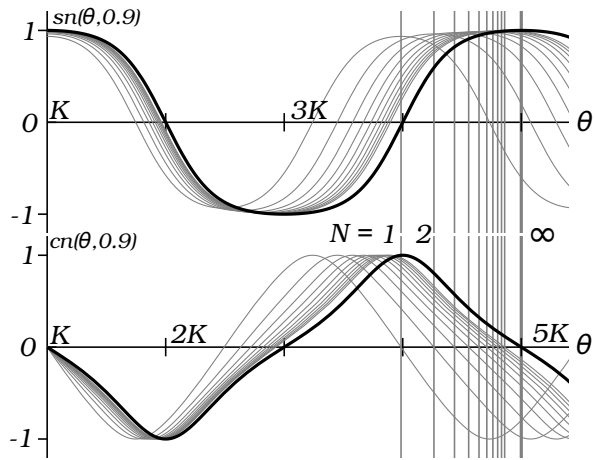} 
  \caption{Approximation Around the Divergence. Approximations of $sn$ and $cn$ for $N\in 1,2,3,\ldots 10$ are shown to 
  approach the exact functions, even at $\alpha = 0.9$. Vertical lines mark the end of one complete period of the $N^{th}$ 
  approximation.}
  \label{fig:SnCnCompare}
\end{center}
\end{figure}

Setting the left hand side of Eqs. 33a-b equal to an $\mathcal{O}(\alpha^N)$ approximation 
allows us to solve for an approximation of both $sn$ and $cn$  \cite{Klee}. Composing approximate trajectories 
$\Psi(\alpha,\phi)$ and time dependence $t(\phi)$ gives parametric function graphs appropriately 
scaled for comparison, as in Fig.~\ref{fig:SnCnCompare}. 

\begin{figure}[h] 
\begin{center}
\includegraphics[scale=.55]{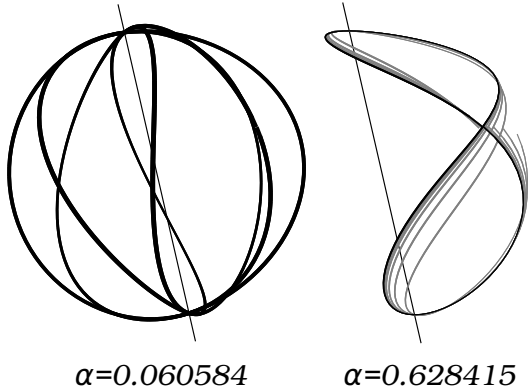} 
  \caption{Approximating Seiffert Spirals. For small $\alpha$, $N=1$ approximations closely follow exact Seiffert spirals (left). 
  Approximate trajectories (gray) for $N=1,2,3,\ldots 6$ approach the exact spiral when $\alpha \approx 0.628$ (right). 
  The $N=7$ approximation (black) appears nearly indistinguishable from the exact spiral.}
  \label{fig:Seiffert}
\end{center}
\end{figure}

Once we extend the approximation to functions such as $cn$ and $sn$, it becomes possible to treat other 
classical motions. We could solve Euler's equations for the free rotational motion of a rigid 
body\cite{Landau}, or describe a photon orbit around a Kerr black hole\cite{Stein}.
Using the inversion relations\cite{Brizard,JacobiErdos}, we could approximate rotational 
motion ($\alpha>1$) of a plane pendulum. We follow Erd\"os\cite{JacobiErdos} by plotting a 
couple of Seiffert spirals\cite{Seiffert}, as in Fig.~\ref{fig:Seiffert}. 

Comparing Figs. ~\ref{fig:TimeEvo},~\ref{fig:SnCnCompare}, and ~\ref{fig:Seiffert} gives an idea of 
limitations encountered when truncating an arbitrary precision result. In a ''small angle'', 
even a simple approximation works well. As $\alpha$ becomes large, more terms in the expansion 
need to be computed. Slow convergence motivates nome expansions\cite{Jacobi}, useful to know 
of, but unnecessary in the present context. 

\subsection{Alternative Approaches}
\subsubsection{Canonical Perturbation Theory}
The Hamiltonian formulation of mechanics also allows one to obtain an 
arbitrary-precision expansion for the phase space trajectory by applying 
a succession of canonical transformations \cite{Lowenstein}. The method 
above is similar in spirit but with a gentle learning curve. 

\subsubsection{Mistaken Expansions}
A great many authors \cite{Landau,Fulcher,Park} recommend 
solving anharmonic oscillations by some variant of the 
Lindstedt-Poincar\'{e} method. This method is error-prone, and usually a 
wrong assumption is made regarding time dependence (cf. Eq. 30), leading 
to something like $\psi_1(\phi) \propto \cos^2(\phi)$ rather than 
$\psi_1(\phi) \propto \cos^4(\phi)$. Taking the wrong $\psi_1(\phi)$, 
it is still possible to compute the correct term of the $f(\alpha)$ series 
expansion, so the mistake often escapes notice. 

\section{Experimental Verification}
The experimental setup, procedure, and analysis for determining the period of a plane pendulum 
are among the most simple and ubiquitous in the physics classroom. A majority of physics students
have completed the basic experiment, while relatively few go on to measure amplitude dependence 
of the period. As is usual in measurement of small perturbations, more stringent precision goals  
require more sophisticated technology. To make fine measurements of the pendulum's motion, we 
need to implement a system with digital data acquisition.

\subsection{Setup and Data Processing}
Experimental systems are available at cost from manufacturers of scientific classroom equipment, but 
a USB mouse device \cite{BallPendulum, OpticalPendulum} provides a thrifty DIY solution, our preference. After modifying the 
mouse into a digital pendulum, we connect it to a Linux work station running the 
X window system \cite{ArchXorg}. The utility program \texttt{xdotool} measures the cursor 
location directly from the computer's desktop environment. Integrating with a \texttt{bash}
script obtains data at nearly millisecond resolution while the pendulum goes through damping 
from maximum amplitude to stop, as in Fig. ~\ref{fig:Data}. 

The oscillation decays through time, only by a small amount per period. Partitioning 
at zero-amplitude intercepts obtains a division of the amplitude vs. time data into 
a number of non-overlapping, nearly-sinusoidal samples of one-period duration. 
Averaging maximum and minimum amplitude, we then associate one-to-one a set of periods 
and a set of amplitudes. Converting amplitude to energy yields a set of period vs. energy 
values. Repeating this process for 100 separate trials,  we obtain a dense sample, 
as in Fig. ~\ref{fig:Data}.

The bulk data obviously shows significant noise, but the sheer number of data points, 
more than $2000$, enables noise reduction by a procedure of binning and averaging.
We set meta-analysis parameters for bin width $\Delta\alpha$ and a minimum energy 
cutoff $\alpha_{min}$ to obtain a more manageable data set, with no apparent 
noise problem.

\subsection{Recursive Data Analysis}
Of course we are not the first to obtain digital pendulum data, or even the first 
to analyze amplitude dependence \cite{EuroMeasure,DigitalPendulum,AnharmMeasure}. 
We fit $K(\alpha)$ using the period as a free parameter and observe good agreement 
over the data range, as in previous investigations. To determine just how well $K(\alpha)$ 
describes the data requires a novel analysis. 

It should be possible to extract expansion coefficients by fitting a cubic function to 
the data, but immediately we encounter a covariance problem. The slope of the data does not
change sign, remains nearly flat. One pass analysis yields inaccurate parameter estimation, 
a wide range of plausible fits. 

\begin{figure}[h] 
\begin{center}
\includegraphics[scale=.22]{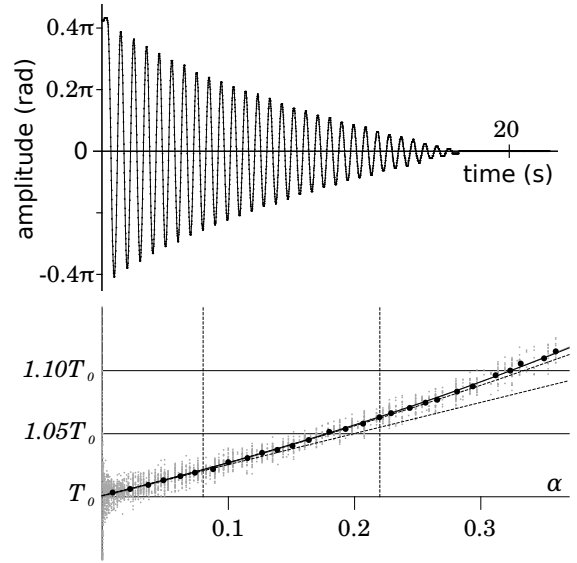} 
  \caption{Measuring $K(\alpha)$. Above: Sample amplitude vs. time data. Below: Period vs. 
	    energy data points in gray are binned and averaged into the black points. Sequential linear  
	    and quadratic fits are depicted as dashed lines, while final cubic fit is a solid line going 
	    through all points. Dashed vertical lines mark the boundaries between data subsets. }
  \label{fig:Data}
\end{center}
\end{figure}

To improve accuracy and precision we take advantage of the data's asymptotic nature by partitioning 
the entire set into three simply connected, disjoint subsets. This introduces two additional analysis priors, 
energy values, $\alpha_{LQ}$ and $\alpha_{QC}$, which demarcate boundaries as in Fig. ~\ref{fig:Data}. 
The fit procedure first determines the period and linear expansion coefficient from linear data, 
subsequently determines the quadratic expansion coefficient from the union of linear and quadratic 
data, finally determines the cubic expansion coefficient from all data. 

In total, the analysis depends on four meta-analysis parameters: $\{\alpha_{min},\Delta\alpha,\alpha_{LQ},\alpha_{QC}\}$. 
To set these values we adopt the following heuristics:
\begin{itemize}
 \item Use as much data as possible. 
 \item Exclude noisy data around $\alpha=0$.
 \item Make the bin width as small as possible.
 \item Capture at least 10 data points per bin.
 \item In the linear range $[0,\alpha_{LQ}]$:
 $$ \frac{2}{\pi}K(\alpha) - (1+\frac{1}{4}\alpha)  < 0.001 \times \frac{2}{\pi}K(\alpha). $$
 \item In the quadratic range $[0,\alpha_{QC}]$:
 $$ \frac{2}{\pi}K(\alpha) - (1+\frac{1}{4}\alpha+\frac{9}{64}\alpha^2)  < 0.001 \times \frac{2}{\pi}K(\alpha). $$
 \item The extracted linear, quadratic, cubic coefficients should have increasing uncertainty. 
 \item Minimize uncertainty where possible. 
\end{itemize}

From these we have initial values $\{\alpha_{min},\Delta\alpha,\alpha_{LQ},\alpha_{QC}\} 
=\{0.003,0.013,0.083,0.21 \}$. Searching around, not too far, 
we find the best fit of Table \ref{tab:params}. Comparison of extracted parameters with coefficients 
of Eq. 6 leads to the humorous conclusion that the Kidd-Fogg formula\textemdash 
though false \textit{de facto}\textemdash also adequately fits the data. That is, 
the cubic fit does not distinguish between competing Eqs.5-6. To decide against 
Kidd-Fogg by data alone requires an experiment with sufficient quality up to and 
beyond $\alpha=0.6$, i.e. $5-10$ expansion coefficients of $K(\alpha)$.

\begin{table}[h]
\begin{center}
\caption{Cubic Best Fit Parameters.} 
\label{tab:params}
$\{\alpha_{min},\Delta\alpha,\alpha_{LQ},\alpha_{QC}\} =\{0.003,0.013,0.08,0.22 \}$
\begin{tabular}{ c | c | c  }
\hline \hline
\;\;\;Expectation\;\;\; & \;\;\;\;\;\; Estimate \;\;\;\;\;\; &  \;\;\;\;\;\;Error\;\;\;\;\;\;  \\
\hline
$\frac{1}{4}\approx 0.2500$ &$0.2463\pm 0.0071$ & $0.52 \sigma \;, \; 1.5\%$  \\
$\frac{9}{64}\approx 0.1406$ &$0.1508\pm 0.0082$ & $1.24 \sigma \; , \; 7.2\%$ \\
$\frac{25}{256}\approx 0.0977$ &$0.1037\pm 0.0126$ & $0.48 \sigma \; , \; 6.2\%$ \\
\end{tabular}
\end{center}
\end{table}

Choosing other vales for $\{\alpha_{min},\Delta\alpha,\alpha_{LQ},\alpha_{QC}\}$ we obtain 
similar best fit parameters, especially when heuristics are nearly obeyed. To facilitate 
comparison of various analyses, archival data and basic tools are available 
online\cite{GITKlee,AURKlee}. 

\section{Quantum Classical Analogy}
The choice of nomenclature in section III suggests a quantum/classical analogy at work. 
The symbol $\Psi$ connotes a quantum wavefunction, but 
above it denotes a C.E.S. Our use of $\Psi$ follows other 
semi-classical works \cite{Harter,RalstonSymp}. In the sequel, we extend the analogy to 
time-independent perturbation theory. 
\subsection{Conservation of Energy}
Whenever we use approximate methods in the analysis of physical systems, 
classical or quantum, we also introduce terms of error at some level of 
precision. For example, approximation of a pendulum's motion may only conserve total energy 
up to some power of $\alpha$. A similar situation often arises in quantum 
mechanics.

We assume a Hamiltonian $H = H_0+\epsilon\;V$ for which the eigenstates $|\psi_n \rangle$ are 
approximately known and non-degenerate,
\begin{subequations}
\begin{align}
H |\psi_n \rangle & =  E_n\;|\psi_n \rangle \\
H_0|\psi_{n,0} \rangle & = E_{n,0}|\psi_{n,0} \rangle.
\end{align}
\end{subequations}
The standard perturbation theory\cite{Peebles} determines  corrections to the zero-order 
wavefunctions and energies. 

To first order, the time-independent Schr\"odinger equation becomes
\begin{equation}
H |\psi_{n,1} \rangle  = ( E_{n,0}+ \epsilon\; E_{n,1})|\psi_{n,1} \rangle +\mathcal{O}(\epsilon^2).
\end{equation}
We make the ansatz
\begin{equation}
|\psi_{n,1} \rangle  = |\psi_{n,0} \rangle + \sum_{i \neq n}\epsilon\;c_{n,i}^1\; |\psi_{i,0} \rangle,
\end{equation}
and solve for
\begin{equation}
c_{n,i}^1 = \frac{\langle \psi_{i,0} |V|\psi_{n,0} \rangle}{E_{n,0}-E_{i,0}}, \;\;\;\;\;
E_{n,1} = \langle \psi_{n,0} |V|\psi_{n,0} \rangle. 
\end{equation}

As with classical perturbation theory, quantum perturbation theory allows iteration to arbitrary 
$N$ \cite{SinghQuant}, where the approximate eigenfunctions 
\begin{equation}
|\psi_{n,N} \rangle  = |\psi_{n,0} \rangle + \sum_{j=1}^N \sum_{i \neq n}\epsilon^j\;c^j_{n,i}\; |\psi_{i,0} \rangle,
\end{equation}
are nearly stationary with regard to energy 
\begin{equation}
H|\psi_{n,N} \rangle  = \sum_{i=0}^N \epsilon^i \; E_{n,i} |\psi_{n,N} \rangle +\mathcal{O}(\epsilon^{N+1}).
\end{equation}
Though again a law of diminishing returns applies to higher order corrections. 

The semiclassical analogy associates conservation of energy with the eigenvalue equation. 
In either theory, iteration of a recursive algorithm changes the shape of a
C.E.S. or a wavefunction such that the perturbed solution becomes 
increasingly precise ( Cf. Fig. ~\ref{fig:QuantPert} ). 

As time evolves the $N^{th}$ classical approximation satisfies
\begin{equation}
\frac{dE}{dt}=0 + \mathcal{O}(\alpha^{N+2}).
\end{equation}
To find an analogous equation in quantum dynamics, we apply an 
infinitesimal time-translation by expanding the Hamiltonian propagator
\begin{eqnarray}
& |\psi_{n,N}(t + \Delta t)\rangle  = e^{-\frac{i \Delta t}{\hbar} H }|\psi_{n,N}(t)\rangle  \\
 & \approx    (1-\frac{i \Delta t}{\hbar} H )|\psi_{n,N}(t)\rangle \nonumber \\
& \approx    (1-\frac{i \Delta t}{\hbar} \sum_{i=0}^N \epsilon^i \; E_{n,i} )|\psi_{n,N}(t)\rangle +\mathcal{O}(\epsilon^{N+1}) \nonumber. 
  \end{eqnarray}
This equation shows that time-evolution acts on the approximate eigenfunctions as a change of complex 
phase, but only to $\mathcal{O}(\epsilon^{N+1})$. Complex phases cancel in expectation products, so 
Eq. 41 implies no worse convergence than 
\begin{eqnarray}
\frac{d\langle H \rangle}{dt} = 0 +  \mathcal{O}(\epsilon^{N+1}).
\end{eqnarray}

As ever, the analogy involves an obvious fallacy: $\alpha$ and $\epsilon$ are 
dimensionless quantities belonging to two separate classes. In the classical theory, we 
suppress dependence on the $\epsilon$ coefficients and implicitly assume that a convergence 
criteria can always be stated as a maximum value for $\alpha$ given an approximation and 
a precision goal. The quantum theory requires quantization of $\alpha$. After more 
exploration and explicit calculation, we hope to gain a better understanding of the 
semiclassical analogy's inner workings. 

\subsection{Quantum Anharmonic Oscillator}
\begin{figure}[h] 
\begin{center}
\includegraphics[scale=.25]{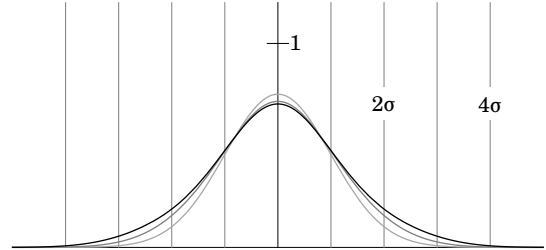} 
  \caption{Perturbed oscillator wavefunction. From light gray to black, the approximate wavefunctions 
  for $N=0,1,2$ with parameter values $\epsilon_1=-2$ and $ h/\lambda_0 = 1$. }
  \label{fig:QuantPert}
\end{center}
\end{figure}

We consider the quantum anharmonic oscillator, with Hamiltonian 
\begin{equation}
H = \frac{\omega_0}{2}(p^2 + q^2) + \frac{\omega_0}{24}\frac{\epsilon_1}{\lambda_{\pi}}\;q^4.
\end{equation}
Using the technique of ladder operators, it is relatively easy to solve for 
energy to $\mathcal{O}(\epsilon)$, 
\begin{subequations}
\begin{align}
E_{n,0} &= \frac{1}{2}(2n+1)\hbar\omega_0,  \\
E_{n,1} &= \frac{1}{32}(2\;n^2+2\;n+1)\hbar\omega_0 ,
\end{align}
\end{subequations}
where $\hbar = h / (2\pi)$ is the reduced Planck's constant and $\epsilon = \epsilon_1 \frac{h}{\lambda_{0}}$. 

Setting equal quantum and classical energies, we see that 
\begin{eqnarray}
\alpha & = \frac{E_{n,0} +\epsilon\;E_{n,1}}{\lambda_{\pi}\;\omega_0}
+\mathcal{O}(\epsilon^2)  \\
&= \frac{1}{2}\frac{h}{\lambda_{0}}(2\;n+1)  \nonumber \\
&+ \frac{\epsilon_1}{32}(\frac{h}{\lambda_{0}})^2 (2\;n^2+2\;n+1)
+\mathcal{O}(\epsilon^2) \nonumber,
\end{eqnarray}
or equivalently, by Eq. 22a, 
\begin{equation}
\lambda(n) = (n +\frac{1}{2})h + \frac{\epsilon_1}{64}\frac{h^2}{\lambda_0} +\mathcal{O}(\epsilon^2). 
\end{equation}
The \textit{quantum conditions} Eqs. 44 \& 45 recall the Sommerfeld-Wilson 
prescription of old quantum mechanics\cite{Tomonaga,ChildBook}
\begin{equation}
\lambda(n) = \oint p \; dq = (n+\delta)\;h,
\end{equation} 
with \textit{Maslov index} 
\begin{equation}
\delta = \frac{1}{2}+ \frac{\epsilon_1}{64}\frac{h}{\lambda_0},
\end{equation}
slightly perturbed from $\delta = 1/2$, the usual value associated with 
harmonic vibrational motion. 

In the case of a ''quantum pendulum'', $\epsilon_1$ can not be made small, so fidelity of
approximate methods depends entirely on the constant $\lambda_{0}$. As $l$ and $m$ become 
increasingly small,  $\lambda_{0}$ approaches $h$. Whenever $h/\lambda_{0}<1/10$ and 
$n < 5$ then $\alpha<1/2$, a first or second approximation adequately describes the 
quantized wavefunctions. The $N=0,1,2$ approximations of the quantum anharmonic 
oscillator's groundstate wavefunction appear in Fig. ~\ref{fig:QuantPert}, with 
extreme expansion parameter $h/\lambda_0=1$. 

\section{Conclusion}
The pendulum takes an eminent place in the physics canon, not only as a 
measurement device but also as an example of anharmonic 
oscillation. The simple harmonic approximation leaves 
open the possibility of spectacular failure because it only 
reliably determines the overall scale of time. Apprehension of the 
dependence on $\theta_0$ or $\alpha$ requires more careful and 
detailed analysis. Here we present a novel algorithm, which solves 
equations of motion and produces the expansion coefficients of $K(\alpha)$. 
Calculations require minimal technical skill and avoid any 
confusing artifice. 

Thrift experiment produces good enough 
data. Taking into account theoretical expectations by writing out 
a list of prior beliefs, we define an analysis where the extracted 
parameters closely match the expected values.  Nothing precludes our
analysis from applying to higher energy motions in the domain 
$\alpha \in [0,1]$. It would be interesting to see how many 
expansion coefficients this method may accurately and 
precisely determine. As many as five, ten?

Perturbation methods extend beyond the important but simple example of 
a plane pendulum. Extending the ansatz Eq. 19 to include 
half-powers of $\alpha$ allows arbitrary precision solution of any one-dimensional, 
power-series potential. This important elaboration leads to applications in 
mathematical biology\cite{Lotka,ArnoldDiffEq}, classical astronomy\cite{KleeClassical}, 
and relativistic astronomy \cite{KleeRelativistic}. In higher dimensional phase space, 
we obtain angular equations of motion along a variety of multidimensional C.E.S. 
Subsequent work will explore the multidimensional generalization. 

The classical/quantum analogy reveals fundamental principles that apply throughout 
physics. Time-independent perturbation theories subject phase space trajectories 
and wavefunctions to perturbative variations. Evolving through time, trajectories 
and wavefunctions meet the expectation that higher precision approximations 
more nearly conserve energy. Exploration of quantum conditions resolves a fallacy 
in the analogy by showing that quantum theory replaces continuous energy parameter 
$\alpha$ with a quantum number $n$. We have yet to find any connections to 
the Matheiu functions\cite{QuantPend}, but speculate of their existence. 

Ultimately we reach a detailed understanding of the plane pendulum and its relation 
to time. The pendulum is a particular anharmonic oscillator, 
with a period that varies slightly as a function of amplitude. On 
the quantum scale atomic oscillations, for example in Cesium-133, provide 
the highest precision time scales. On the astronomical scale, we also measure 
long times, even for the practical purpose of keeping a calendar. Seconds are 
useful units for time, but physics also needs nanoseconds and years.
Wherever we find an oscillation, usually anharmonicity is not too far behind, 
leading to unexpected behaviors such as precession.  

\section*{Acknowledgments}
The author gratefully acknowledges dissertation committee members, William Harter, Salvador Barraza-Lopez,
and Daniel Kennefick for helpful discussions, comments on earlier drafts of 
the paper; also, Wolfdieter Lang and many other volunteer editors for their work 
on the OEIS. This work was supported in part by a Doctoral Fellowship awarded by the 
University of Arkansas. 

\bibliographystyle{unsrtnat}
\bibliography{Pendulum}

\end{document}